\documentstyle[12pt]{article}

\newcommand{\lsim}{\stackrel{<}{\sim}} 

\def\phibar{{\bar \phi}}
\def\psibar{{\bar \psi}}

\def\lattint{\int {d^4k \over (2\pi)^4}}
\def\lattintt{\int {d^3k \over (2\pi)^3}}

\begin{document}

\begin{titlepage}
\begin{flushright}
\hfil                 UCSD/PTH 92-22\\
\hfil June 1992 \\
\end{flushright}
\begin{center}

{\LARGE\bf The Finite Temperature Phase Diagram \\
\vspace{0.3cm}
of a U(1) Higgs-Yukawa Model}

\vspace{2.0cm}
{\large Karl Jansen}\\
\vspace{0.1cm}
University of California at San Diego\\
Department of Physics-0319\\
La Jolla, CA~92093-0319\\
USA  \\
\vspace{3.0cm}
\end{center}

\abstract{
The finite temperature phase diagram of a U(1) Higgs-Yukawa model
at a finite value of the scalar self coupling $\lambda$ is
investigated by means of a large-$N_f$ calculation and numerical
simulations. The phase diagram is similar to the one at zero temperature
and shows a ferromagnetic, two symmetric and an antiferromagnetic
phase. However, the phase transition lines are shifted to larger values of the
Yukawa coupling demonstrating the occurence of the finite temperature
symmetry restoration.}
\vfill
\end{titlepage}

\section{Introduction}

The finite temperature electroweak phase transition plays an important
role in scenarios of the early universe
which assume that at a high enough temperature the universe was in a hot
symmetric state \cite{brand}. During its expansion the universe cooled
down and at a critical temperature passed through the electroweak phase
transition breaking the symmetry spontaneously.
That the symmetry should be restored at high enough temperatures was
already predicted in the work of
Kirzhnits and Linde \cite{kili}.
A subsequent analysis \cite{doja,weinberg} within the
framework of large $N$ and perturbative approximations confirmed
the symmetry restoration picture. To test the validity of these proximations,
it would be useful to investigate
the phenomenon of symmetry restoration also
non-perturbatively. Moreover, questions like the order of the phase
transition or the value of the critical temperature $T_c$ can not be
answered reliably in perturbation theory alone. Nonperturbative
lattice simulations may provide more insight into the nature of the
electroweak phase transition. However,
the inclusion of all degrees of freedom is
too demanding with present day computers
and one has to restrict oneself to subsystems of the standard model
like gauge-Higgs systems \cite{gahi} or pure scalar theories.
In particular in the O(4) symmetric $\phi^4$ theory
the critical temperature can be estimated assuming that
the gauge and fermionic degrees of freedom can be neglected.
As the gauge coupling is rather weak this seems to be a reasonable
approximation and one finds
$T_c \approx 350GeV$ \cite{jase}.

In this letter I want to include the fermions by studying a
U(1) Higgs-Yukawa system in the large fermion number ($N_f$)
approximation in combination with
lattice simulations.
Higgs-Yukawa models on the lattice
have been the object of numerous analytical and
numerical studies in the last years \cite{goltermann}.
They are expected to provide
insight into non-perturbative properties of the Standard model.
They revealed, for example, a surprisingly complex phase diagram.
For a finite value of
the scalar self coupling $\lambda$ four different phases were found at
zero temperature \cite{UCSD,hajapd}.
These phases are a ferromagnetic phase (FM), where the symmetry is
spontaneously broken, two symmetric phases (SYM), one at weak
and the other at strong Yukawa coupling and an antiferromagnetic
(AFM) phase where the staggered magnetization is non-zero
with two regions, separated by a first order phase
transition line. For an infinite scalar self coupling $\lambda=\infty$
even an additional phase was found, a ferrimagnetic (FI) phase where the
magnetization as well as the staggered magnetization is non-zero
\cite{bodepd}.

At small scalar self coupling $\lambda$ the system can be analyzed by
analytical means like large-$N_f$, perturbative or mean-field
approximations. The
non-perturbative simulation results are in good agreement with these
analytical calculations \cite{hajapd}. In particular it could be
confirmed that the phase transition from the ferromagnetic to the
symmetric phase at weak Yukawa coupling, which is relevant for the
standard model, is second order with the critical behaviour of the
Gaussian fixed point. This means the triviality of the Higgs Yukawa
models. Of course, the model is plaqued by the appearance of
doubler fermions so
that one can not expect quantitative results from these studies.
Nevertheless, Higgs-Yukawa models are expected to
describe qualitative features
correctly like
the order of the phase transitions and the
question of symmetry restoration which I want to study here.

\section{The model and the techniques}

\subsection{Zero temperature}

The lattice action for the $U(1)$ chiral
invariant Higgs-Yukawa model at zero temperature
is defined as
\begin{equation}
S = S_f + S_H  ~.
\label{eq:action}
\end{equation}
The fermion part of the action $S_f$ is given by
\begin{equation}
S_f = \sum_{x,z} \psibar_i(x) M(x,z) \psi_i(z)~~, ~~
i = 1,2,...,N_f/2 ~~.
\label{eq:F-act}
\end{equation}
In eqn. (\ref{eq:F-act}) the fermion matrix may be written as
\begin{equation}
M(x,z) = \sum_{\mu} \gamma_\mu\left[\delta_{x+\mu,z}-\delta_{x-\mu,z}\right]
+ y\left[\phi_1(x)+i\gamma_5\phi_2(x)\right]\delta_{x,z} ~,
\label{eq:F-matrix}
\end{equation}
where $\gamma_\mu, \gamma_5$ are the Hermitian Dirac matrices and $y$ stands
for
the Yukawa coupling.
The scalar part of the action $S_H$ in eqn. (\ref{eq:action})
is given by
\begin{eqnarray}
S_H &=& -\kappa\sum_{x,\mu} \phi_a(x)\left[\phi_a(x+\mu)+\phi_a(x-\mu)\right]
+ \sum_x \phi_a(x)^2  \nonumber \\
&+&  \sum_x\lambda \left[\phi_a^2(x)-1\right]^2 ~,~~
a = 1, 2 ~~.
\label{eqnarray:H-act}
\end{eqnarray}
The scalar fields $\phi$ are complex, $\lambda$ is the scalar self
coupling and $\kappa$ is the hopping parameter which is related to the
mass parameter $m_c$ in the continuum formulation,
$m_c^2 = (1-2\lambda-8\kappa)/\kappa$.
The sums in eqs.(\ref{eq:F-act}) and (\ref{eqnarray:H-act})
run over the lattice size
$\Lambda =L_TL_s^3$.

The model defined in eqns.(\ref{eq:action}-\ref{eqnarray:H-act})
can be solved in the large fermion number ($N_f$) limit.
It is convenient \cite{hajapd} to consider a modified form of the scalar action
\begin{eqnarray}
S_H &=& -\kappa_N \sum_{x,\mu}
\varphi_a(x)\left[\varphi_a(x+\mu)+\varphi_a(x-\mu)
\right] + \sum_x \varphi_a(x)^2 \nonumber \\
&+& {\lambda_N} \left(\varphi_a^2(x)-N_f\right)^2 ~.
\label{eqnarray:F-act2}
\end{eqnarray}
The usual lattice action  eqns.(\ref{eq:action}-\ref{eqnarray:H-act})
is obtained by identifying
\begin{equation}
\kappa_N = C^2 \kappa~, \
\lambda_N = C^4 \lambda  ~,\
y_N = C y~,
\label{eq:factor}
\end{equation}
where the factor $C$ satisfies the equation
\begin{equation}
C^4 - (1-2\lambda_N N_f) C^2 - 2\lambda_N = 0
\label{eq:C}
\end{equation}
The scalar field $\varphi_a(x)$ is related to the original field $\phi_a(x)$
by
\begin{equation}
\varphi_a(x) = \phi_a(x)/C~.
\end{equation}
In the $1/N_f$ expansion  the couplings
$\tilde y_N =\sqrt {N_f}y_N$, $\tilde \lambda_N=N_f\lambda_N$
are kept fixed to be $O(1)$. As $N_f \to \infty$ the relations
in eqn. (\ref{eq:factor}) simplify, giving
\begin{equation}
\kappa = {\kappa_N \over {1-2\tilde\lambda_N }}~, \
\lambda N_f = {\tilde\lambda_N \over {(1-2\tilde\lambda_N)^2}} ~,\
y \sqrt{N_f} = { \tilde y_N \over {\sqrt{1-2\tilde\lambda_N}}} ~, \ \ \ \
{\tilde\lambda_N < {1\over 2}}~. \
\label{eq:factor2}
\end{equation}
In the large $N_f$ limit the constant mode of the scalar field dominates
the path integral which suggests the Ansatz
\begin{equation}
\varphi_1(x) = \sqrt{N_f} \left[(a + (-1)^{\sum_\mu x_\mu} b \right]\;\;,
\varphi_2(x) = 0 ~,
\end{equation}
where $\sqrt{N_f}a$ and $\sqrt{N_f}b$  correspond to the
magnetization and staggered magnetization, respectively. The effective
potential at leading order is
\begin{equation}
{1\over N_f} V_{eff}(a,b) = -8\kappa_N(a^2-b^2)+(a^2+b^2)+{\tilde\lambda_N}
\left((a^2+b^2-1)^2+4a^2b^2\right) - 2 I(a,b,0)
\label{eq:effpot}
\end{equation}
where the zero temperature lattice integral is
\begin{equation}
 I(a,b,0) =\lattint \log\left[ \sum_\mu
\sin^2k_\mu + \tilde y^2_N(a^2-b^2)\right]~.
\label{eq:I0}
\end{equation}

\subsection{Finite temperature}

A finite temperature on the lattice can be realized by a finite
extension $L_T$ in the temporal direction. The other directions on the
lattice are kept large enough so that finite size effects are negligible.
This leads to the usual finite temperature geometry of the lattice
$L_s^3L_T$ with $L_s \gg L_T$.
The physical temperature is given in terms of the lattice spacing $a$
\begin{equation}
T=1/L_T a ~.
\label{eq:phystemp}
\end{equation}
Varying the physical temperature
can be achieved by changing $L_T$.
Symmetry restoration on the lattice can then be detected in the following way:
Assume that there is a phase transition from a symmetric to a broken
phase at zero temperature. If now by heating the system, i.e. decreasing
$L_T$, this phase transition shifts into the broken phase,
we have found the symmetry restoration.
Fixing the parameters of the theory, characterized by, for example, the
renormalized scalar self coupling $\lambda_r$ and the renormalized
Yukawa coupling $y_r$,
to be in the broken phase at zero temperature,
the system passes the finite temperature phase transition
at some $L_T$ corresponding to a critical physical temperature $T_c$ and
for values of $T>T_c$ the system is in the symmetric phase.

The lattice action of the finite temperature U(1) Higgs-Yukawa model is
the same as in eqs.(\ref{eq:action}-\ref{eqnarray:H-act}).
The only change is that
now the sums run over the finite temperature lattice with $L_s \gg L_T$. The
steps of the large-$N_f$ approximation are therefore very similar to the
zero temperature case. The only change is that
at finite temperature the expression of $I(a,b,0)$, eq.(\ref{eq:I0}), has to
be replaced by
\begin{equation}
I(a,b,L_T) = \frac{1}{L_T}\sum_n \lattintt \log
\left[ k_4^2 + \sum_i\sin^2 k_i +\tilde
y^2_N(a^2-b^2)\right]\; ;i=1,2,3
\label{eq:IT2}
\end{equation}
where $k_4 = 2\pi(n+1)/L_T, n=0,...,L_T-1$. Note that in contrast to the
continuum the sum over $n$ is finite and can not be
done explicitely.

Replacing $I(a,b,0)$ by $I(a,b,T)$ in (\ref{eq:effpot}) gives the finite
temperature effective potential.
One may obtain different phases
correponding to the locations of the minima of $V_{eff}$.

\noindent (1)Symmetric (SYM) solution: There is a single minmum at $a=b=0$.

\noindent (2)Ferromagnetic (FM) solution: There is a minimum at $a\ne 0, b=0$

\noindent (3)Antiferromagnetic (AFM) solution: The minimum is at $a=0, b\ne0$

\noindent (4)Ferrimagnetic (FI) solution: The minimum is at $a\ne0, b\ne 0$.
It can be shown that this solution does not exist for small values of
the scalar self coupling $\lambda \lsim 1$.

To determine the phase structure I have calculated the integral
(\ref{eq:IT2}) which is needed for the effctive potential
(\ref{eq:effpot})
numerically. I used mainly finite lattice sums as approximations to
save computertime. However, I
checked for various points that the lattice integrals give compatible results.

As described in \cite{affleck,UCSD,hajapd}, for
large $y$ values another type of large $N_f$ expansion is
possible. Here one keeps the Yukawa coupling $y_N \sim O(\sqrt{N_f})$ and
$\kappa_N \sim O(1/\sqrt{N_f})$. As a result of this large-$N_f$ expansion
one obtains an effective action which is the $XY$-model in four
dimensions
\begin{equation}
S_{eff} = -\kappa_{eff}
\sum_{x,\mu}\sigma_a(x)\left[\sigma_a(x+\mu)+\sigma_a(x-\mu)\right]
\end{equation}
\noindent where the fields $\sigma_a(x)$ have unit length,
$\sigma_a(x)\sigma_a(x)=1$.
The effective hopping parameter
$\kappa_{eff}$ is given by
\begin{equation}
\kappa_{eff} = \kappa_N\varphi_0^2+{N_f \over 2y_N^2\varphi_0^2}~.
\label{eq:XY}
\end{equation}

At finite temperature the steps of the calculation are the same with
$\kappa_{eff}$ replaced by the one of the XY-model at finite temperature,
$\kappa_{eff}(T)$.
I determined $\kappa_c(T)$ for the XY-model
on lattices with $N_T =6$ and $N_T =8$ in time direction
by a numerical simulation using the cluster algorithm. I
find the phase transition still to be of second order and
$\kappa_c(T)$ shifted only slighly to larger $\kappa$-values.
This shift is less than $1\%$ similar to case of the O(4) model
\cite{jase}.
Therefore the phase transition lines from the large
$N_f$ expansion in the strong Yukawa coupling region show only a tiny shift.

\subsection{Numerical simulation}

I have performed  numerical simulations at $\lambda = 0.0156, N_f = 2$.
Note that these parameters are the same as the ones used in Fig.2a of
ref.\cite{hajapd} for zero temperature.
The Hybrid Monte Carlo method \cite{Kennedy} was used for the dynamical fermion
simulations. Each molecular
dynamics trajectory consists of $10$ steps with step size chosen such that
the acceptance rate is around $80\%$.
As a check I did simulations at $\lambda=0.0156$ for $N_f=10$
and found agreement with
the large-$N_f$ expansion.
To decide the order of the phase transition,
I looked for  hysteresis effects in the thermocycles.
For each data point
in the thermocycle about 50
trajectories are used as warmup and 100-200 trajectories are used
in the measurement.

The order parameters to detect the phasetransitions have been
the magnetization $v$ defined as
\begin{equation}
v = <\sqrt{\phibar_a^2}>, \ \   \phibar_a = {1\over L^4}\sum_x \phi_a(x)
\end{equation}
and the staggered magnetization $v_{st}$ given by
\begin{equation}
v_{st} = <\sqrt{\phibar^2_{st,a}}>, \ \  \phibar_{st,a} = {1\over L^4} \sum_x
(-1)^{\sum_\mu x_\mu} \phi_a(x) ~,
\end{equation}
where $L$ is the linear size of the lattice.
The measurements are done on $8^32$ and a few on $10^32$ lattices.
Although these lattices are certainly not sufficient to distinguish second
order
and weakly first order transitions, the
combination and the agreement
of the numerical and analytical results give a reliable
determinination of the order of the phase transitions.

\section{Discussion and conclusion}

The results of the large $N_f$ and Monte Carlo calculations are shown in
fig.1 for the Yukawa coupling $y \le 1.5$. I have left out the results
for the strong Yukawa region as they are indistinguishable from the zero
temperature case. I checked explicitely also in this region
that the Monte Carlo data agree with the large $N_f$ predictions.

In the figure the large-$N_f$ results are shown as lines, where
full lines indicate second order and dashed lines first order phase
transitions. The Monte Carlo results for the phase transitions are
exhibited as circles where open symbols mean first order and full symbols
second order phase transitions. I find agreement between the
Monte Carlo results and the theoretical prediction from the large $N_f$
calculation not only for the position but also for the order of the
phase transitions.
There occur three different phases, a ferromagnetic (FM), a
symmetric (SYM) and an antiferromagnetic (AFM) phase. As mentioned
earlier, at large values of the Yukawa coupling a second symmetric phase
appears, which is not shown in fig.1.
The dotted lines in fig.1 indicate the phase diagram at
zero temperature as found in \cite{hajapd}. Although the structure and
the order of the phase transitions remain the same at finite
temperature, a clear shift to larger values of the Yukawa coupling is
seen.

Let me concentrate on the phase transition between the ferromagnetic
(FM) and the symmetric (SYM) phases which is relevant for the standard
model. As discussed above, the observed shift to larger values of the
Yukawa coupling indicates the expected symmetry restoration:
A point (denoted as the star in fig.1) chosen to be in
the broken (FM) phase at zero temperature will be found in the symmetry
restored phase at finite temperature.
It is noteworthy that for the moderate temperatures that
could be tested here the phasetransition is still second order and in
the domain of attraction of the Gaussian fixed point. I conclude that
symmetry restoration is not only an effect of a large $N$ or
pertubative apparoximation but survives also when non-perturbative
methods are applied.

\noindent{\bf Acknowledgement}

I want to thank Y.Shen for several useful discussions.
This work is supported by DOE grant at UC San Diego (DE-FG-03-90ER40546).


\pagebreak

{\bf Fig.1} The finite temperature phase diagram at
$\lambda = 0.0156$ and $N_f = 2$. The MC data are indicated by circles
where the solid symbols denote second and the open symbols first order
phase transitions. The solid and dashed lines are the
results from the $1/N_f$ expansions, where the solid lines represent
second order and the dashed line first order phase transitions. The
phases are: (FM) ferromagnetic, (SYM) symmetric and (AFM)
antiferromagnetic.
The dotted lines indicate the zero temperature phase
diagram \protect{\cite{hajapd}}. Note that the phase transitions for
finite temperature are shifted to larger values of the Yukawa coupling.
This shows the expected finite temperature symmetry restoration as
indicated by the starred point which is in the broken phase at zero and
in the symmetric phase at finite temperature.

\end{document}